\input{aipcheck}

\documentclass[
    ,final            
  ]
  {aipproc}

\layoutstyle{8x11single}

\def\be{\begin{equation}}
\def\ee{\end{equation}}
\def\ba{\begin{eqnarray}}
\def\ea{\end{eqnarray}}
\newcommand{\psidag}{\psi^{\dag}}

\begin{document}

\title{Renormalization group analysis of reggeon field theory: flow equations}


\author{Jochen Bartels}{
  address={II. Institut f\"{u}r Theoretische Physik, Universit\"{a}t Hamburg, Luruper Chaussee 149, D-22761 Hamburg, Germany}
}

\author{Carlos Contreras}{
  address={Dep. de Fisica, Uni. Tecnica Federico Santa Maria, Aveda.Espana 1680, Casilla 110-V, Valparaiso, Chile}
}

\author{G. P. Vacca}{
  address={INFN Sezione di Bologna, via Irnerio 46, I-40126 Bologna, Italy}
}

\begin{abstract}
Can large distance high energy QCD be described by Reggeon Field Theory as an effective emergent theory? We start to investigate the issue employing functional renormalisation group techniques.
\end{abstract}

\maketitle

\section{Introduction}
\label{sec:intr}
To find a bridge between short and long distances physics in the high energy QCD remains 
a topic of high interest. At small transverse distances where perturbation theory can be applied QCD predicts the BFKL Pomeron~\cite{BFKL}. 
More general, the BFKL Pomeron can be seen as the basic building block of QCD reggeon field \cite{Gribov:1968fc} theory in which the reggeized gluon 
is the basic field and the Pomeron is generated as a composite state of two or more reggeized gluons. Within this perturbative QCD reggeon field theory, the BFKL Pomeron 
is expected to receive corrections, have self interactions\cite{Bartels:1994jj} leading to Pomeron loops etc.    
On the other hand, high energy scattering of hadrons strongly depends upon large transverse distances where perturbation theory is not applicable. As the most promising 
theoretical concept, again Regge theory ~\cite{Gribov:1968fg,AB,Migdal:1973gz} can be used, but the parameters have to be 
taken from data. Most prominent and phenomenologically successful examples include the Regge pole model of Donnachie and Landshoff \cite{Donnachie:2013xia},
the reggeon field theory model of Kaidalov et al \cite{Kaidalov:1986tu}, and, more recently, the two models based upon summation of Pomeron diagrams of the Tel-Aviv \cite{Gotsman:2008tr} and the Durham group \cite{Ryskin:2011qe}.   

We therefore see strong evidence that in both regions  - short and long transverse distances - we have the same structure: reggeon field (RFT) which lives in one time (rapidity) and two space dimensions (transverse distances) (these variables are conjugate to reggeon energy $\omega$ and and transverse momenta $k$):

\be
S(\psi,\psidag) = \int d\tau d^2x {\cal{L}}(\psi(\tau,x) \psidag(\tau,x)),
\ee

In its simplest version, RFT is based upon Regge poles. What is different in the two regions are the parameters of the Regge poles (intercepts, slopes) and their interactions. Whereas at short distances they can be computed form QCD perturbation
 theory \footnote{At short distancees perturbative QCD is described by the BFKL Pomeron; if 
one imposes boundaries in the IR region, the BFKL Pomeron turns into a sequence of Regge poles with intercepts greater than one},  the parameters describing the long distance region, so far, have to taken from data. 

The theoretical challenge then is to find a bridge between the two regions.   
Whereas in recent years much work has been done in analyzing the short distance region,
less is known for the long distance region \cite{Gribov:1968fg,AB,Migdal:1973gz}.
Here we want to address the following question: can RFT be considered a useful effective description which can be linked to QCD starting from the knowledge of its formulation (fields and symmetries)?

The tool we are going to use is the renormalisation group technique in an essentially wilsonian form, i.e. we shall make use of the effective average action (EAA) description~\cite{Wetterich},
which allows to study the change of the generator of the proper vertices of the theory as one integrates over the ultraviolet field modes.
The analysis of the flow equation for the effective average action has successfully been applied to numerous problems in statistical mechanics, in particle physics, and in quantum gravity.

The EEA, $\Gamma_k[\phi]$, belongs to a space of functionals which depends on the field $\phi$ and symmetry content of a theory as well as on the momentum space infrared cutoff $k$ which controls the range of modes which are integrated out. For $k>0$ the infrared region is regulated by some regulator operator $R_k$ which is associated to a quadratic form.
In the limit $k\to 0$ one finds the full effective action of the theory.
In general the EAA is a non local functional, in the sense that cannot be written in terms of a local lagrangian, exactly as its standard effective action counterpart. Nevertheless many properties of the dynamical flow can be studied using some simplifications, i.e. choosing a truncation which consists in projecting the functional on a subspace. Several classifications of such projections have been proposed. One of the most popular is based on a derivative expansion, which starts from the so called local potential approximation (LPA), where apart from a simple kinetic term describing the propagation of the fields, one allows for a pure local potential term  $V_k$ in the lagrangian.

The flow equation~\cite{Wetterich} reads in general 
\be
k \partial_k \Gamma[\phi]=\frac{1}{2} {\rm Tr}\left[ \left(\Gamma^{(2)}[\phi]+R_k\right)^{-1} k \partial_k  R_k \right]
\label{flow}
\ee
and allows to study two important features of the dynamics: first one may ask which are the functionals which are invariant under the flow (the fixed point of the flow) and how the flow behaves close to it (the critical exponents predicting a universal behavior), second one can study the flow from some bare condition close to the UV region towards the IR region and investigate the approximated form of the effective action, depending on the different regions (phases) one starts from. 
The first kind of analysis has to be done in terms of dimensionless quantities, keeping in mind that many physical aspects have to deal with dimensionful ones.
Let us stress that a fixed point can be realised in both the UV or IR regions and that there might exist a particular flow which connects two such points.
In this language a theory is renormalizable (finite and predictive) if there exist a UV fixed point with a finite number of relevant directions (eigenvectors of the associated linearized  flow with positive critical exponents), since a flow starting close to such a fixed point towards the IR will span a finite dimensional subspace, any point on which will be fixed by a finite number of measurements.
In this framework one may introduce a concept of emergence of an effective theory from a microscopic "more" fundamental one. 
Starting from the fields of the fundamental d.o.f. one may see that a convenient description arises performing a change of field variables, with new symmetries for the new degrees of freedom (d.o.f.). This may happen at some point of the microscopic wilsonian flow which will correspond also to a point of the space of the emergent effective theory. At this stage the flow towards the IR will be conveniently governed by the universal properties of the new emergent theory.
    
As discussed above, our ultimate goal is the understanding of the transition from small to large distances in relation to high energy QCD processes. 
A reasonable first step is the analysis of the extreme limit of large transverse distances and high energies. 
In a second step one will hope to address the most difficult part, the transition from one limit to the other. 

We therefore start by using the framework of RFT and address the question 
whether it may provide a good description in the limit of asymptotic large rapidity and transverse distances. We first analyse the fixed point structure of the Wilsonian flow for the EAA,
together with its associated critical exponents.
We shall employ a rather drastic LPA truncation, eventually slightly correcting it with the LPA', which consists in adding to it an approximate knowledge of the anomalous dimensions. 
Furthermore, in this first step we use, as a further approximation, a polynomial expansion for the potential.

In the following we present a brief outline of our calculations and discuss the first results. 

\section{RFT model}
As anticipated, we shall work in a truncation at the lowest order in the derivative expansion, so that the effective action in terms of the pomeron field $\psi$ reads
\begin{equation}
\label{effaction}
\Gamma_k[\psi^{\dag},\psi]  =\int \!  \, \mathrm{d}^D x \,  \mathrm{d} \tau
 \left( Z_k( \frac{1}{2} \psidag\partial_{\tau}^{\leftrightarrow} \psi - \alpha'_k \psidag \nabla^{2} \psi)  + V_k[\psi^{\dag},\psi] \right),
\end{equation}
where $\tau$ is the rapidity, the dimension $d$ of the transverse space will be mainly specialized to $D=2$, and the potential $V_k$ has the following general properties:

\noindent (1) $V_k$ is symmetric under the interchange $\psi \leftrightarrow \psidag$,
(2) for real values of $\psi$ and $\psidag$, the real part of $V_k$ is symmetric under $\psi \to -\psi$, $\psidag \to -\psidag$, the imaginary part odd.

In order to tackle in the simplest possible way the problem we shall employ a polynomial expansion. This can be done conveniently around some stationary point of the potential, which can be seen in perturbative terms as a vacuum state.
One can consider both an expansion around a zero field (trivial) or around a non zero field (non trivial). Let us stress that in both cases such an expansion will be valid only within some radius of convergence.
In the former case one can write  
\ba
\label{potential}
V[\psi^{\dag},\psi]  =
- \mu \psi^{\dag} \psi + i\lambda  \psi^{\dag}(\psi^{\dag}+\psi)\psi +g  (\psidag \psi)^{2} +g' \psidag ({\psidag}^2 + \psi^2) \psi \nonumber\\
+ i\lambda_5 {\psidag}^2 \left(\psidag + \psi \right) \psi^2 +i\lambda'_5 \psidag \left({\psidag}^3+ \psi^3 \right) \psi +...
\ea
Choosing the Pomeron trajectory function $\alpha(t=-q^2) = \alpha(0) - \alpha' q^2$
we see the relation between the 'mass' parameter  $\mu$ and the intercept $\mu=\alpha(0)-1$.
We note here that the theory defined by this potential has a non Hermitian Hamiltonian (but PT symmetric ~\cite{BV}).

The flow of the potential can therefore be extracted in the constant field (both in rapidity and transverse space) approximation. We shall employ a cutoff operator whose Fourier representation $R_k(\omega,q^2)=Z_{k} \alpha'_{k} (k^{2}-q^{2}) \Theta(k^{2}-q^{2})$, so that the quantum modes are regulated using the so called optimised cutoff in transverse space and a "mass" cutoff in rapidity space. 
To move to dimensionless quantities we shall use $[x]=k^{-1}$ and $[\tau]=E^{-1}$ so that $[\psi] = [\psidag] = k^{D/2}$ and $[\alpha']= E k^{-2}$ and
$\tilde{\psi} = Z_{k}^{\frac{1}{2}} k^{-D/2}\psi$,  $\tilde{V_k}= \frac{V_k}{\alpha'_k k^{D+2}}$.

We shall not give here the detailed expressions for the PDE which describes the flow of $\tilde{V_k}$ 
nor the set of coupled ODEs which describe, for the polynomial approximation, the flow of the couplings such as $\mu$, $\lambda$, $\cdots$, but simply describe few of the results we obtain.
         
As one interesting result, we find numerical evidence for the existence of a fixed point in the multi-dimensional space of coupling constants,
which has one relevant direction (i.e. one negative eigenvalue, all other directions with positive eigenvalues), independen of the order of the polynomial.
We show here the simplest case of an effective average potential which includes only terms of interactions up to cubic order: the vector field plot (Fig.~1) shows a non trivial fixed point which,
if one neglects the anomalous dimensions, is located at $(\mu^* = 0.11111, \,\, \lambda^*=\pm 1.0503)$.
\begin{figure}
  \includegraphics[height=.3\textheight]{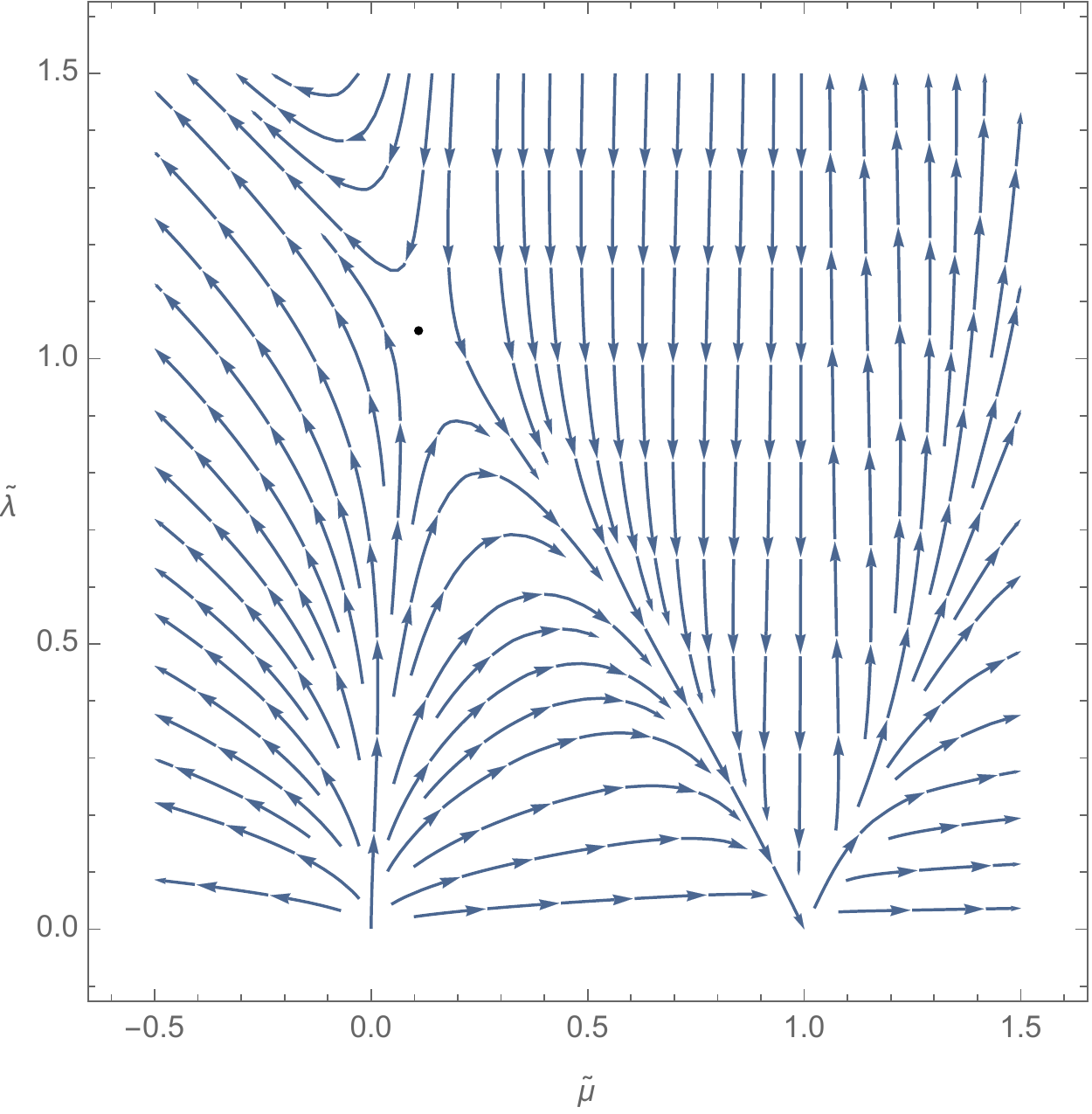}
  \caption{Fixed point structure and vector field for the system of ODE for the cubic truncation ($\mu$,$\lambda$ couplings). The arrows point to the IR direction.}
\end{figure}
The linearized RG flow at this  fixed point is governed by the  the properties of a stability matrix, which has 
the following eigenvalues ($\theta$) and eigenvectors ($v$):
\begin{equation}
 \left(
  \begin{array}{cccccc}
   \mu^* & \lambda^ * & \theta_1 & \theta_2 & v_1 &v_2 \\
  0.00 & 0.00 & -2.00 & -1.00 & (-1,0)&(0,-1) \\ 
  0.11 & +1.05 & +2.39 & -1.89 & (-0.108,-0.994)&(-0.779, +0.674) \\
  0.11 & -1.05 & +2.39 & -1.89 & ( +0.108,-0.994)&(-0.779,-0.674) \\
  \end{array}
\right)
\end{equation}
We note that we have studied also higher order truncations, and these stability properties (one negative eigenvalue, all other directions with positive eigenvalues) persist.
In particular for polynomials of order 3,4,5 we find for the negative eigenvalue respectively (-1.89,-1.69,-1.45).
Similar results are obtained also analysing the flow of $V_k$ from a polynomial expansion around a non "trivial vacuum". We note in this case a very strong improvement on the convergence with the increase of the order of the polynomial.

As our main result  we have found that there exist a non trivial interacting fixed point (for dimensionless quantities) with one single relevant direction associated to it. Moereover, the $\mu$-value associated with the fixed point is positive, i.e. the corresponding intercept is greater than one. The latter matches the fact that in the UV region (short distance region) the BFKL Pomeron predicts an intercept greater than one.

These results allow for two possible interpretations:\\
(i) the fixpoint is seen is a UV fixed point (of the emergent theory, so that it is a "would be" FP) and if the flow of the fundamental QCD theory in the ultraviolet region places, in the new description, the "bare" emergent theory in some region around this fixed point, the flow will be governed by the corresponding universal behavior. One relevant direction means that with one IR measure one could fix such a RFT description.\\
(ii) the fixpoint is interpreted as a IR fixed point.  
The fact that the relevant direction drives away from it could be avoided only if - for some reason to be explained-  the fundamental theory in the ultraviolet region would be fine-tuned  such that the "bare" RFT action lies on a 'critical surface': In the multi-dimensional space of  coupling constants this critical surface is defined to be orthogonal to the one-dimensional 
relevant direction, and for all trajectories inside this surface the fixpoint is attractive in the IR region. A solution inside this critcal surface maybe related to the scaling solutions of  'critical reggeon field theory' which have been studied in \cite{AB,Migdal:1973gz} in the vicinity of four transverse dimensions. One might guess that for such a scenario a possible reason could be the unitarity of the fundamental theory (QCD).

Finally we would like to stress that, so far, our discussion has been formulated in terms of 
dimensionless parameters. Physical quantities, i.e. the value of the Pomeron intercept,
carry dimension. In order to determine, for the two scenarios described above, physical 
observables, we have to solve the flow equations (\ref{flow}) for the Pomeron Green's function.
Work along these line is in progress and will be published elsewehere.\\ \\
\begin{theacknowledgments}
JB thanks the INFN sezione di Bologna for the hospitality. CC gratefully acknowledges the  financial support from proy. FONDECYT 113099, DGIP/USM 11.13.12 and from the DAAD. 
\end{theacknowledgments}

\bibliographystyle{aipproc}

\end{document}